\documentstyle[11pt,epsfig]{article}
\title{
Transversity Distribution Functions in the Valon Model
 }

\author{ Z.Alizadeh Yazdi$^{(a)}$\footnote{zahra.alizadehyazdi@stu.um.ac.ir}, F.Taghavi-Shahri$^{(a)}$\footnote{f\_taghavi@ipm.ir} , F.Arash $^{(b)}$\footnote{farash@cic.aut.ac.ir} and M.E.Zomorrodian $^{(a)}$\footnote{ zomorrod@ferdowsi.um.ac.ir}
\\
$^{(a)}$ Department of Physics, Ferdowsi University of Mashhad,
P.O. Box 1436, Mashhad, Iran\\
$^{(b)}$ Physics Department, Tafresh University, Tafresh, Iran\\
 }
\date{\today}
\begin{document}
\maketitle

\begin{abstract}
We use the valon model to  calculate the  transversity distribution functions , inside the Nucleon. Transversity distributions indicate
the probability to find  partons with spin aligned (anti- aligned)
to the transversely polarized nucleon. The results are in good
agreements with all available experimental data and also global
fits.

\end{abstract}

\section{Introduction}

The  nucleon "spin crisis" is still one of the most fundamental
problems in high energy spin physics. Results of   Deep Inelastic
Scattering (DIS) experiments suggest that just $30\%$  of the spin
of the proton is carried by the intrinsic spin of its quark
constituents. This discovery has challenged our understanding
about the internal structure of the proton. Therefore many
theoretical and experimental studies have been conducted to
investigate and
understand the role of spin in the proton's internal structure.\\
The key question is how the spin of the nucleon is shared among
its constituent quarks and gluons. That is, the determination and
understanding of the shape of quarks and gluon
spin distribution functions have become an important task. \\
In general, there are three collinear parton distribution functions: the unpolarized parton distribution
functions (PDFs), the longitudinally  polarized distribution
functions (PPDFs) and the transversity distributions  .
 They are defined as follows:
 if we show the number density of quarks with helicity $\pm 1$ inside a
 positive hadron with $q_{\pm }(x,Q^2)$, then we have:
\begin{equation}\label{pdf}
q(x,Q^2)=q_{+ }(x,Q^2)+ q_{-}(x,Q^2)
\end{equation}
\begin{equation}\label{deltaq}
\Delta q(x,Q^2)=q_{+ }(x,Q^2)- q_{-}(x,Q^2)
\end{equation}
where $q(x,Q^2)$  is the probability of finding a parton with
fraction $x$ of parent hadron momentum and $\Delta q(x,Q^2)$
represents the probability of finding a polarized parton with
fraction $x$ of parent hadron momentum and spin align/anti-align
to hadron's spin. It measures  the net helicity of
partons in a longitudinally polarized hadron.\\
The third parton distributions are  transversity distribution functions. They have a simple meaning too: In a transversely
polarized hadron, transversity distribution is denoted by $\Delta_T q(x,Q^2)$ and
represents the number density of partons with momentum fraction $x$
and polarization parallel to that of the hadron minus the number
density of partons with the same momentum fraction and
antiparallel spin direction:
\begin{equation}
 \Delta_T q(x,Q^2)=q_{\uparrow }(x,Q^2)-q_{\downarrow}(x,Q^2)
\end{equation}
Historically they were first introduced in 1970's by Ralston and
Soper \cite{Ralston} and rediscovered by Artru and Mekhfi \cite{artru} in  the
beginning of 90's and their
 QCD evolution studied by Jaffe and Ji  \cite{Jaffe}.\\
Since  $\Delta_T q(x,Q^2)$ is a chirally-odd  quantity, it can not
be probed in the cleanest hard process, DIS.  It can only be
accessed in  process where it couples to another chirall-odd
quantity. As such, $\Delta_T q(x,Q^2)$ can be  measured  in  hard reactions such as
  semi-inclusive leptoproduction or in the Drell-Yan di-muon production.
Measuring the transverse polarization of partons are the goal
 of  experiments such as COMPASS, HERMES, RHIC and
 SMC Collaborations \cite{COMPASS,HERMES,group}.
These measurements can teach us  about
 the transvesity distribution and the transverse motion of quarks and thus the role that their
 orbital angular momentum play in the structure of proton and
 fragmentation processes.\\
Calculation of  transversity distribution functions, using some
phenomenology is an active task in spin physics \cite{soliton,Bag
model,Anselmino2009,Anselmino2013}. We  intend to do the same and
calculate transversity  distribution using the Valon model.  The
valon model is a phenomenological model originally proposed by R.
C. Hwa, \cite{Hwa1980} in early 80's. It was improved later by Hwa
\cite{Hwa1995} and Others \cite{Hwa2002,Arash2003, Arash prc2003}
and extended to the polarized cases \cite{Arash2007, Arash2008,
Arash2010}. In this model a hadron is viewed as three (two)
constituent quark-like objects, called valons. Each valon is
defined to be a dressed valence quark with its own cloud of sea
quarks and gluons. The dressing processes are described by QCD.
The structure of a valon is resolved at high $Q^2$. At low $Q^2$,
a valon  behaves as constituent quark of the hadron. In this model
the recombination of partons into hadrons is a two stage process:
in the first step the partons emit and absorb gluons in the
process of the evolution of the quark- gluon cloud and become
"valons"; then these valons recombine into hadron. The model
describes the  un-polarized and
 polarized nucleon structure  rather well \cite{Arash prc2003,Arash2010}.\\
 In the present paper we apply  the valon concept
to the transverse polarization and calculate the transversity distribution functions. The paper
is organized as follows. In Section 2 we review
 the valon model for calculating the polarized parton distribution
 functions(PPDFs). Then in  Section 3  we utilize it to calculate   the transversity  distribution. Our conclusions are given in Section 4.

\section{Polarized parton distribution functions in the  valon model }

In the valon representation of hadrons the polarized parton
distribution in a polarized hadron is given by:
\begin{equation}\label{ppdf}
\delta q_{{i}}^{\it{h}}(x,Q^2)=\sum \int_{x}^{1}
\frac{dy}{y}\delta G_{\it{valon}}^{h}(y)  \delta
q_{{i}}^{\it{valon}}(\frac{x}{y},Q^2)
\end{equation}
where $\delta G_{\it{valon}}^{h}(y)$ is the helicity distribution
of the valon in the hosting hadron i.e (probability of finding the
polarized valon inside the polarized hadron). Here we study the
internal structure of proton, so we have to use the polarized
valon distributions inside proton.  $\delta G_{\it{valon}}^{p}(y)$
is related to unpolarized valon distribution, $G_{j}^{p}(y)$ by:
\begin{equation}\label{pvalon}
\delta G_{j}^{p}(y) = \delta F_{j}(y) G_{j}^{p}(y)
 =N_{j}y^{\alpha_{j}}(1-y)^{\beta_{j}}(1+ a_{j} y^{0.5} + b_j y +c_j y^{1.5} +d_j y^2)
\end{equation}
 where $j$
refers to U and D type valons \cite{Hwa1980,Arash prc2003}.
Polarized valon distributions are determined by a phenomenological
argument
 \cite{Arash2007}. The parameters
in Eq. (\ref{pvalon}) are summarized in Table (1) and $\delta
G_{\it{valon}}^{p}(y)$ are plotted in Figure (1). The term $
\delta q_{{i}}^{\it{valon}}(x/y,Q^2)$ for $h=p$ in Eq. (\ref{ppdf})
is the polarized parton distribution inside a valon. Their
evolution are governed by the DGLAP equations
\cite{DGLAP1,DGLAP2,DGLAP3}. Finally, the polarized proton
structure functions are obtained via a convolution integral as
follows:
\begin{equation}\label{gone}
g^{p}_{1}(x,Q^{2})=\sum_{\it{valon}}\int_{x}^{1}\frac{dy}{y}
\delta G_{\it{valon}}^{p}(y) g^{\it{valon}}_{1}(\frac{x}{y},Q^{2})
\end{equation}
where $g^{\it{valon}}_{1}(\frac{x}{y}, Q^{2})$ is the polarized
structure function of the valon. The details of the actual
calculations are given in \cite{Arash2007,Arash2008,Arash2010}.

\begin{figure}[htp]\label{udvalon}
\centerline{\begin{tabular}{cc} 
\includegraphics[width=8.5 cm]{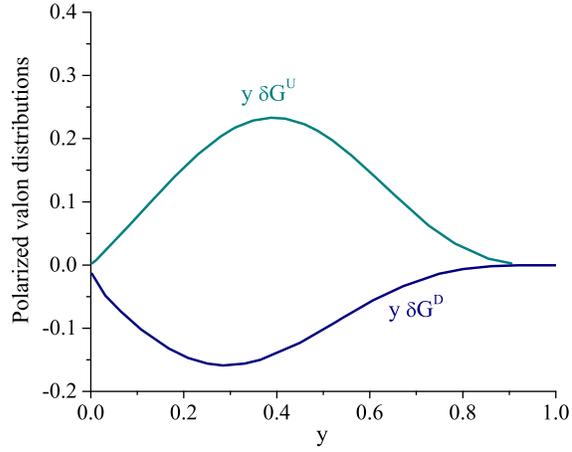}
\end{tabular}}
 \caption{\footnotesize
(color online) Polarized valon distributions for U and D valons inside the
proton.   }

\end{figure}

\begin{table}\label{parameters}
\begin{center}
\begin{tabular}{|c|c|c|c|c|c|c|c|c|}
\hline $valon (j)$ & $N_{j}$ & $\alpha_{j}$ &
$\beta{_j} $ & $a_j $ & $b_j $ & $c_j $ & $d_j$ \\
\hline
$U$ & 3.44 & 0.33 & 3.58 & -2.47 & 5.07 & -1.859 & 2.780 \\
$D$ & -0.568 & -0.374 & 4.142 & -2.844 & 11.695 & -10.096 & 14.47
\\
\hline
\end{tabular}
\caption{Numerical values of the parameters in Eq. (\ref{pvalon}) for polarized valon distributions inside proton. }

\end{center}
\end{table}

\section{Transversity Distribution Functions in the valon model}

We now follow the same procedure as in Section 2, to calculate the
transversity distribution functions of partons in the proton. For
the transversely polarized proton,  Eq. (\ref{ppdf}) reads as:
\begin{equation}\label{tmds}
\Delta_{T} q_{{i}}^{\it{p}}(x,Q^2)=\sum_{valons} \int_{x}^{1}
\frac{dy}{y}\Delta_T G_{\it{valon}}^{p}(y)  \Delta_{T}
q_{{i}}^{\it{valon}}(\frac{x}{y},Q^2)
\end{equation}
where $ \Delta_T G _{valon}^{p} (y) $ is the transverse valon
distribution functions describing the probability of finding a
valon with spin aligned or anti-aligned with the transversly
polarized proton. In fact, $ \Delta_T G _{valon}^{p} (y) $ is
identical to $\delta G_{\it{valon}}^{p}(y) $ in the longitudinal
case. This is so,  because we know  that in the non-relativistic
limit of the quark motion, the PPDFs and transversity distribution
would be identical, since the rotations and Euclidean boosts
commute and a series of boosts and rotation can  convert a
longitudinal polarized proton into a transversely polarized one
with an infinite momentum  \cite{Anselmino2009,ts1}. The only
difference between the transversity distributions and PPDFs
reflects the relativistic character of quark motion in the proton
and shows up in the splitting functions and DGLAP equations.
Consequently,  here we set $ \Delta_T G _{valon}^{p} (y) = \delta
G_{\it{valon}}^{p}(y) $. Also notice that $
\Delta_{T}q_{i}^{valon}(\frac{x}{y},Q^2) $  in Eq. (\ref{tmds})
are the transversity distribution functions
in the valon. They  can be calculated using the DGLAP evolution equations, as described bellow.\\
In the Mellin space, transversity distribution functions are given by:
\begin{equation}\label{one}
\Delta_{T}q_{\pm}(n)=\Delta_{T}q(n)\pm\Delta_{T}\overline{q}(n)
\end{equation}
where $\Delta_{T}q_{\pm}(n)$  are the Singlet and Non-Singlet
transversity distribution functions of partons. The first moment
(n=1) of transversity distribution refers to the proton's tensor
charge \cite{Jaffe1,Jaffe2,soffer}. Their  DGLAP evolution
equations are \cite{Ellis}:

\begin{equation}\label{two}
\frac{d}{d \ln
Q^{2}}\Delta_{T}q_{-}(n,Q^{2})=\Delta_{T}\gamma_{qq,-}(n,\alpha_{s}(Q^{2}))
\Delta_{T}q_{-}(n,Q^{2})
\end{equation}
\begin{equation}\label{three}
\frac{d}{d \ln
Q^{2}}\Delta_{T}q_{+}(n,Q^{2})=\Delta_{T}\gamma_{qq,+}(n,\alpha_{s}(Q^{2}))
\Delta_{T}q_{+}(n,Q^{2})
\end{equation}
The solution of the DGLAP evolution equations in the Mellin space
at NLO approximation are \cite{Reya}:

\begin{eqnarray}\label{dglaptmd}
\Delta_{T}q_{\pm}(n,Q^{2}) & = & \{1+\frac{\alpha_{s}(Q_{0}^{2})-\alpha_{s}(Q^{2})}{\pi \beta_{0}}[\Delta_{T}\gamma_{qq,\pm}^{(1)}(n)-\frac{\beta_{1}}{2\beta_{0}}\Delta_{T}\gamma_{qq}^{(0)}(n)]\} \nonumber \\
&&
\times(\frac{\alpha_{s}(Q^{2})}{\alpha_{s}(Q_{0}^{2})})^{\frac{-2\Delta_{T}\gamma_{qq}^{(0)}(n)}{\beta_{0}}}\Delta_{T}q_{\pm}(n,Q_{0}^{2}),
\end{eqnarray}
\\
In the above equation, $ \Delta_{T}q_{\pm}(n,Q_{0}^{2})$ are the
initial input densities. They are determined   by a
phenomenological argument in the valon model.
$\Delta_{T}\gamma_{qq,\pm}^{(0)}(n)$ and
$\Delta_{T}\gamma_{qq,\pm}^{(1)}(n)$ are the usual anomalous
dimensions and are given in Appendix A.\\
In the following, first we solve the DGLAP evolution equations for a valon. This
will give  transversity distribution functions in each valon. We then use them in the convolution
integral, Eq. (\ref{tmds}) to obtain transversity distribution functions in the proton.
In doing so, we adopt the $\overline{MS}$ scheme with
$\Lambda_{QCD}=0.22$ $GeV$ and $Q_{0}^{2}=0.283$ $GeV^2$. This value of  $Q_{0}^{2}$ corresponds to
 a distance of $0.36 fm$ which is roughly equal to  or slightly less than the radius of a
 valon. It may be objected that such distances are
probably too large for a meaningful pure perturbative treatment.
We note that valon structure function has the property that it
becomes $\delta(z -1)$ as $Q^2$ is extrapolated to $Q_{0}^{2}$ ,
which is beyond the region of validity. This mathematical boundary
condition signifies that the internal structure of a  valon cannot
be resolved at $Q_{0}^{2}$ in the NLO approximation. Consequently,
when this property is applied to Eq. (\ref{tmds}), the structure
function of the nucleon becomes directly related to $x \delta_T
G^{valon}$ at those values of $Q_{0}^{2}$. Furthermore, as noted
in \cite{Arash prc2003}, we have checked that when $Q^2$
approaches $Q_{0}^{2}$, the quark moments approach to unity and
gluon moments go to zero. From the theoretical standpoint, both
$\Lambda_{QCD}$ and $Q_{0}^{2}$ depend on the order of the
moments, but here, we have assumed that they are independent of
moment order. In this way, we have introduced some degree of
approximation to the $Q^2$ evolution of the valence and sea
quarks. However, on one hand there are other contributions like
target-mass effects, which add uncertainties to the theoretical
predictions of perturbative QCD, while on the other hand since we
are dealing with the valons, there is no experimental data to
invalidate moment order independent of $\Lambda_{QCD}$. Therefore
we led to choose our initial input densities at $Q_{0}^{2}$
 to be $\delta(z - 1) $,  leading to:
\begin{equation}
\Delta_{T} q_{+ }(z,Q_0^{2})= \Delta_{T} q_{- }(z,Q_0^{2})=
\delta(z -1)
\end{equation}
Thus,  their moments are
\begin{equation}
\Delta_{T} q_{+}(n,Q_0^{2})= \Delta_{T} q_{-}(n,Q_0^{2})=
\int^{1}_{0} z^{n-1} \delta(z - 1)dz=1
\end{equation}
It is also interesting to note that our selected value for $Q_0^2$
is very close to the transition region reported by the CLAS
Collaboration for  the behavior  of the first moment of the proton
structure
function  around $Q^2=0.3\  GeV^2$ \cite{CLAS}.\\
The moments of valence quark transversity distribution is now
easily obtained from the solution of DGLAP evolution equations,
Eq. (\ref{dglaptmd}), in  Mellin space; as they are  shown in figure (2).
Finding the transversity distribution functions in a valon, using
Eq. (\ref{dglaptmd}), is now reduced to an inverse Mellin transformation. This
enable us  with the help of  Eq. (\ref{tmds}) to obtain   $ x
\Delta_{T}u(x)$ and $x \Delta_{T}d(x)$ as a function of x. They
are shown in figure (3) for a number of $Q^2$-values.

\begin{figure}[htp]\label{first}
\centerline{\begin{tabular}{cc}
\includegraphics[width=8.5cm]{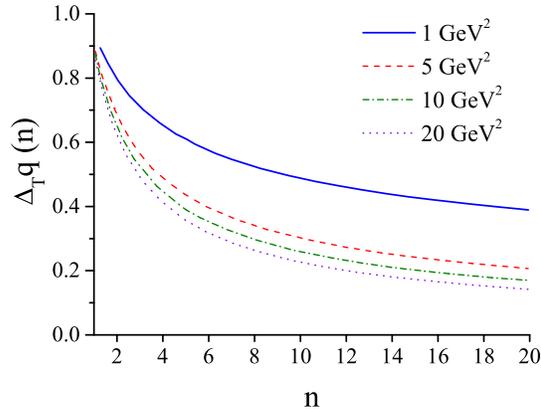}
\end{tabular}}
 \caption{\footnotesize (color online)  $ \Delta_{T}q(n)$ as a function of n in
different ranges of $Q^2$
 .}
\end{figure}
\begin{figure}[htp]\label{xvtq}
\centerline{\begin{tabular}{cc}
\includegraphics[width=8.5cm]{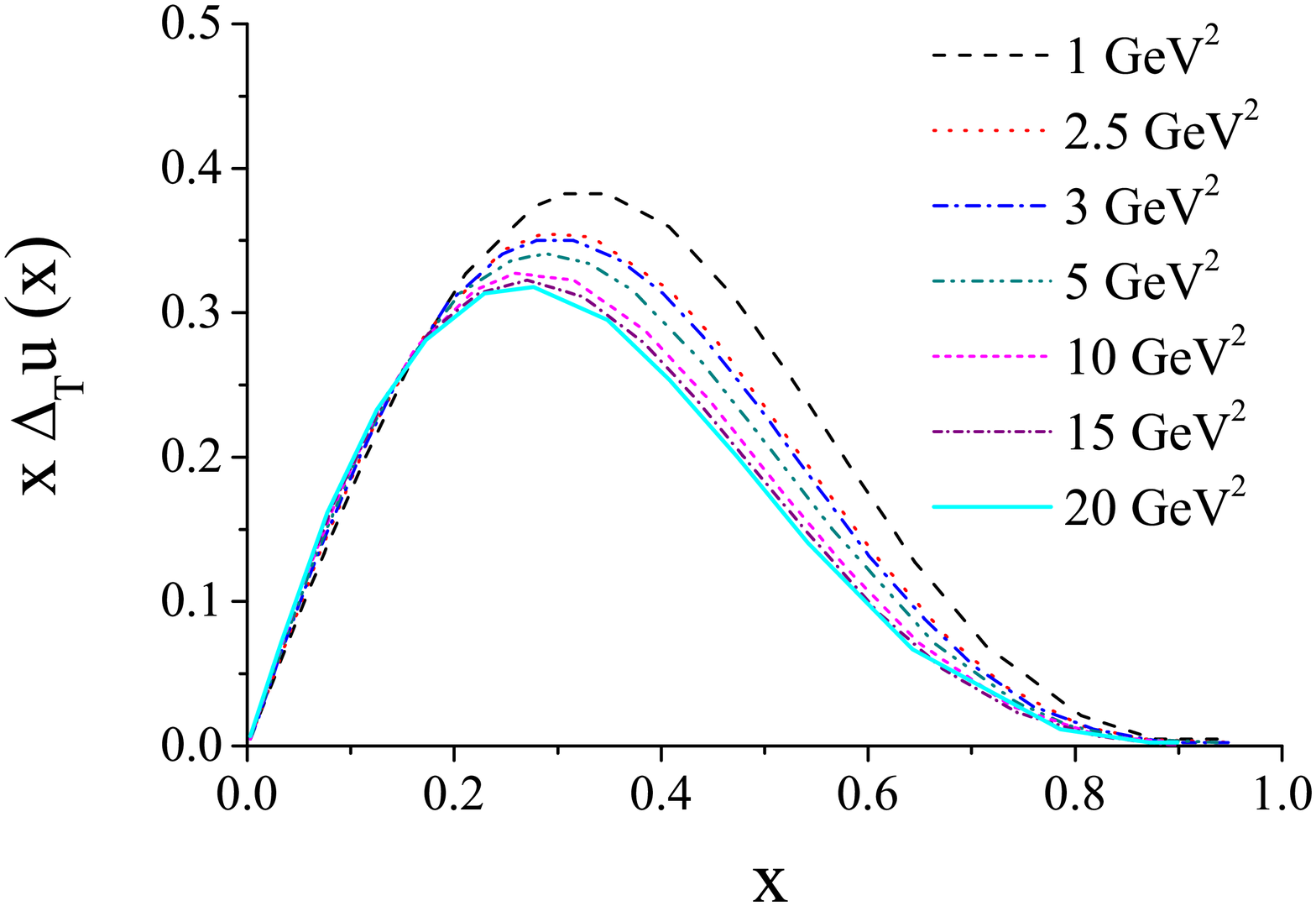}
\includegraphics[width=8.5cm]{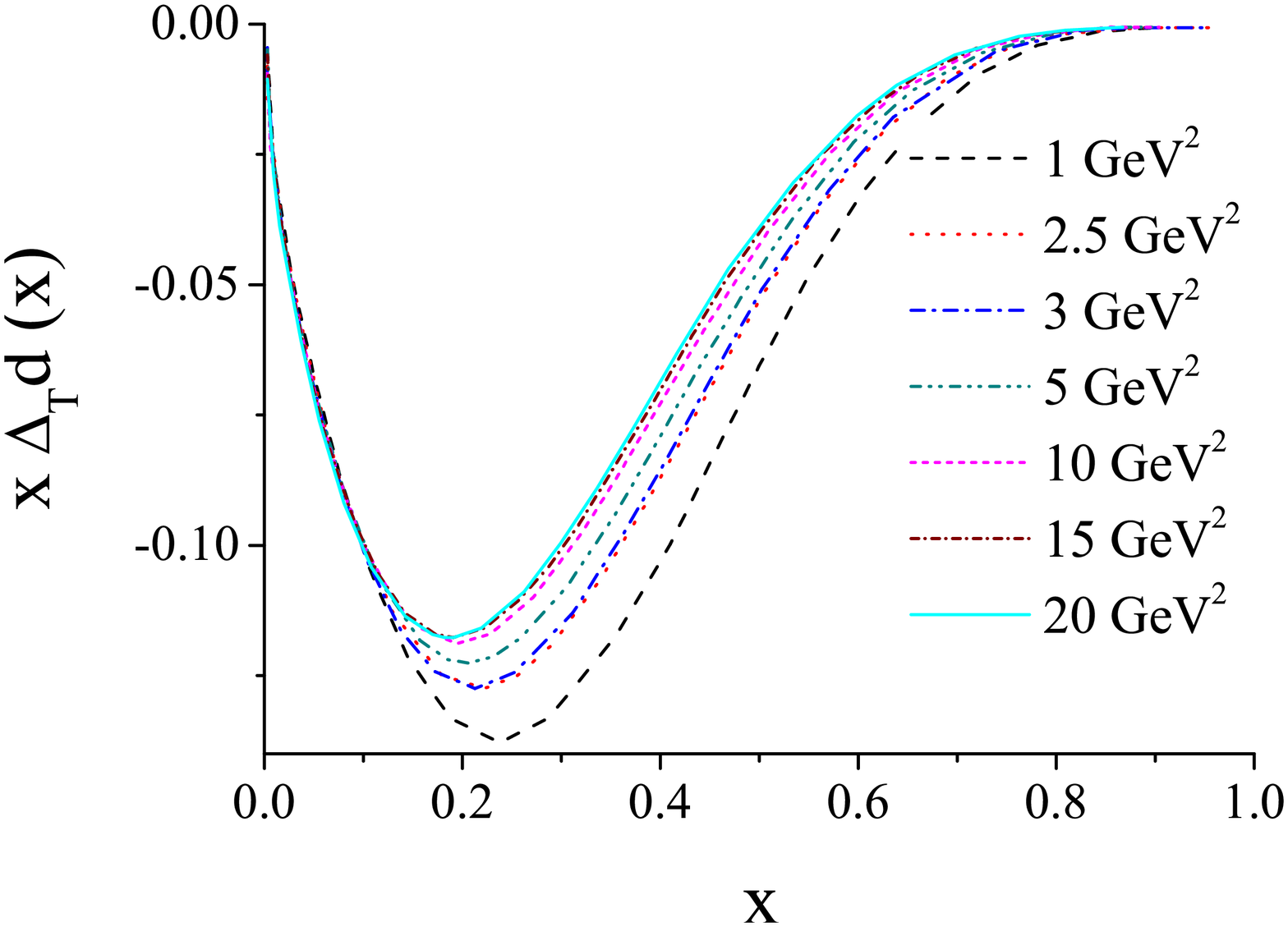}
\end{tabular}}
 \caption{\footnotesize (color online)  $ x \Delta_{T}u (x)$ and  $x \Delta_{T}d (x)$ as a function of x
 for
different ranges of $Q^2$
 .}
\end{figure}
It is common to write the transversity distribution functions as:
\begin{equation}
\Delta_{T}q(x)= \int \Delta_{T}q(x,k_{\bot})\  d^2 k_{\bot}
\end{equation}
where $\Delta_{T}q(x,k_{\bot})$ are the un-integrated  transversity distribution functions. We
 assume that $ k_{\bot}$ dependence of transversity distributions are  factorized in a Gaussian form:
\begin{equation}
\Delta_{T}q(x,k_{\bot})=
\Delta_{T}q(x)\frac{e^{\frac{-k_{\bot}^{2}}{\langle
k_{\bot}^{2}\rangle}}}{\pi\langle k_{\bot}^{2}\rangle},
\end{equation}
where $ \Delta_{T}q(x) $ is  transverse distribution function and
the average values of $ k_{\bot} $ is taken from  SIDIS cross
section data \cite{Anselmino2005,Anselmino2007}, to be
\begin{equation}
\langle k_{\bot}^{2}\rangle = 0.25 \  GeV^{2}
\end{equation}
In figure (4), We show our results for  the transversity distribution function of the valence u quark
, $ x\Delta_{T}u(x,Q^2) $. It is
compared with Anselmino 2008 and Soffer 's global fits  at
$Q^{2}=2.4 \ GeV^{2}$ \cite{Anselmino2009,soffer2}. We also show the result for $x \Delta_{T}u(x,k_{\bot})$ distribution at $x=0.1$ in the right panel of figure (4) .
The same plot is given for d valence quark in figure (5).
Figure (6) shows a more recent global fit results \cite{Anselmino2013} as compared to our analysis.

\begin{figure}[htp]\label{xuvt}
\centerline{\begin{tabular}{cc}
\includegraphics[width=8.5cm]{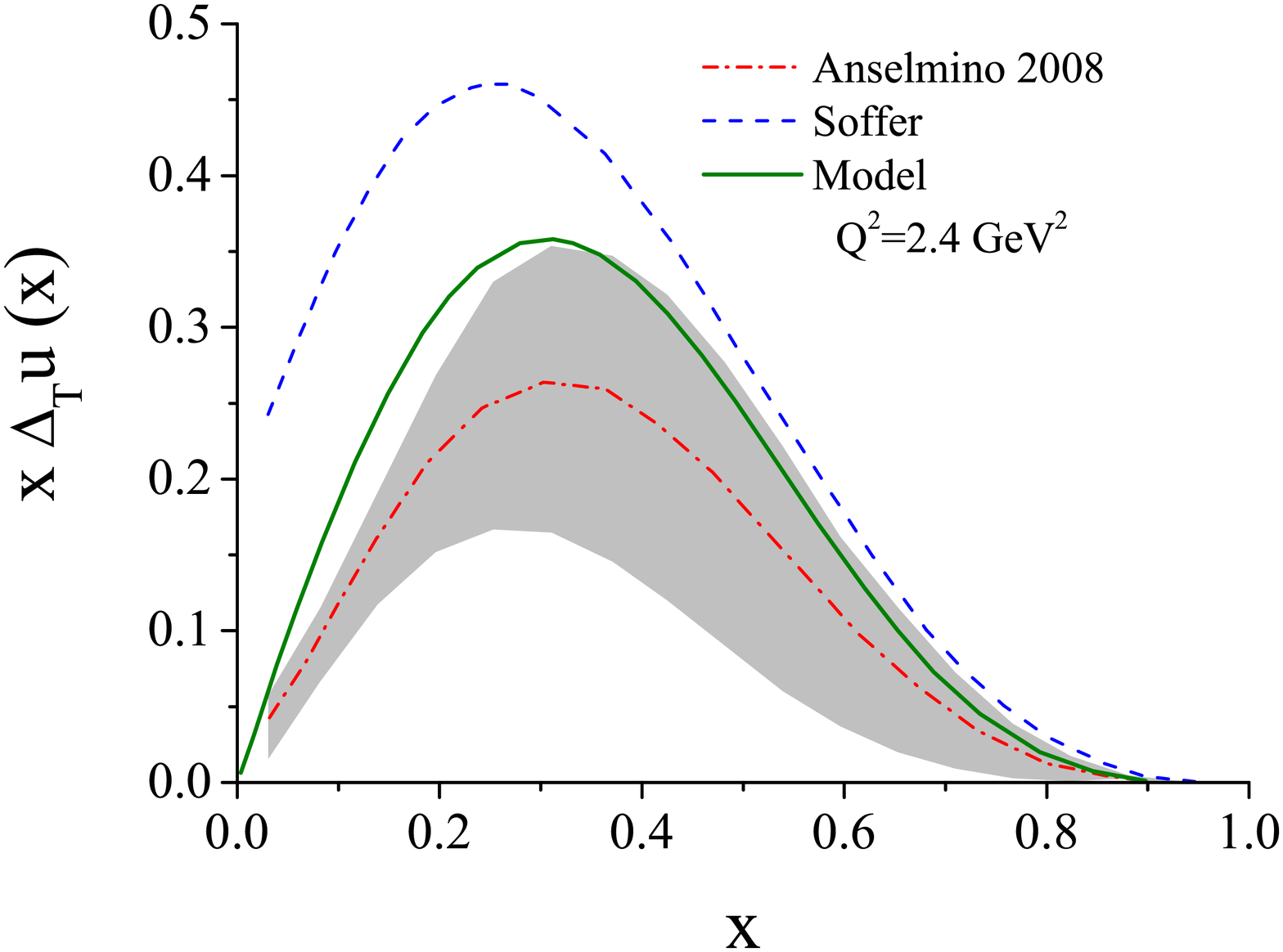}
\includegraphics[width=8.5cm]{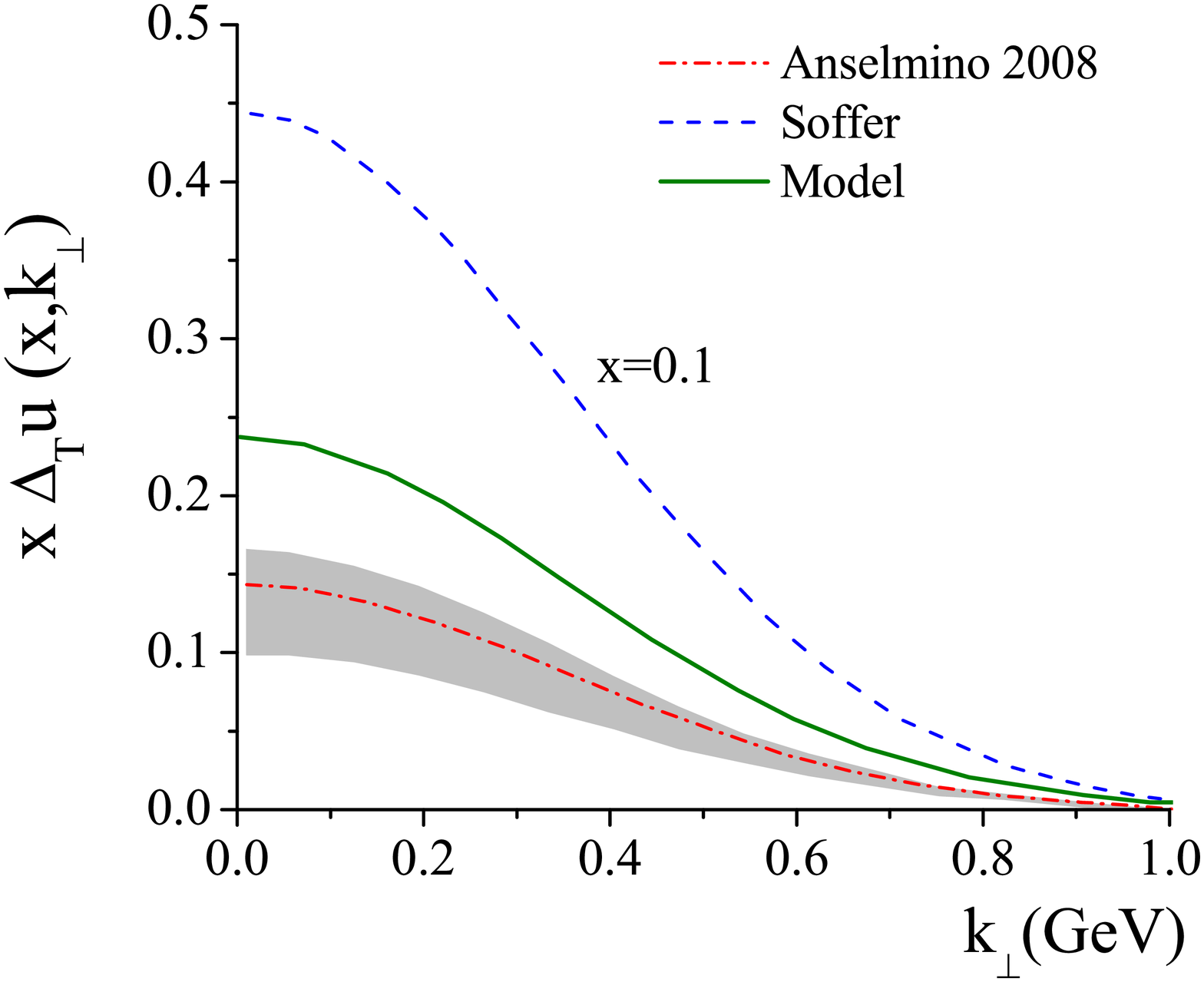}
\end{tabular}}
 \caption{\footnotesize (color online)  The transversity distribution function for  valence u  quark calculated by our model
as a function of x and  $k_\bot$  at $Q^{2}=2.4 \ GeV^{2}$ . They are compared with those from Soffer and Anselmino
global fits \cite{Anselmino2009,soffer2}.
 }
\end{figure}

\begin{figure}[htp]\label{xdvt}
\centerline{\begin{tabular}{cc}
\includegraphics[width=8.5cm]{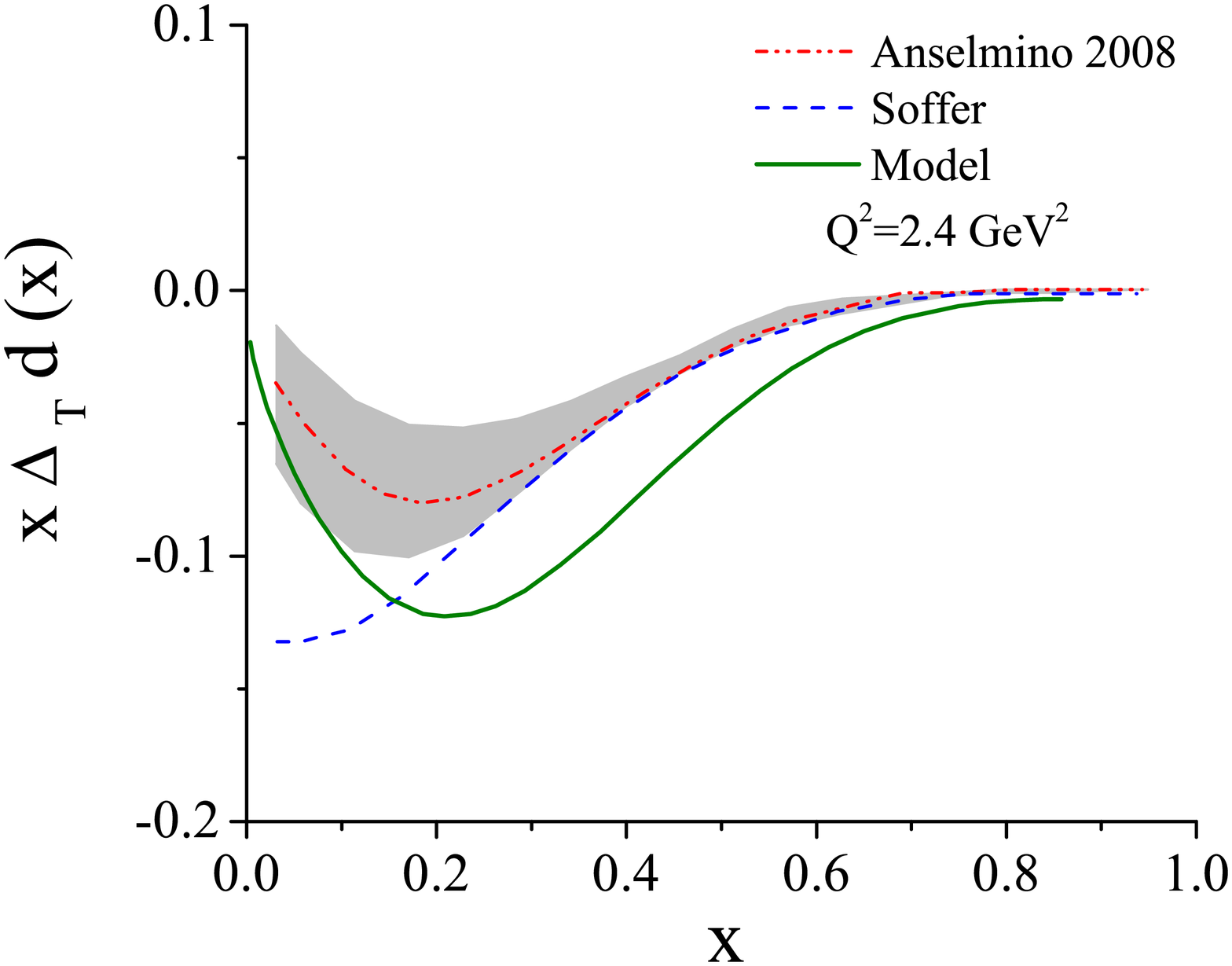}
\includegraphics[width=8.5cm]{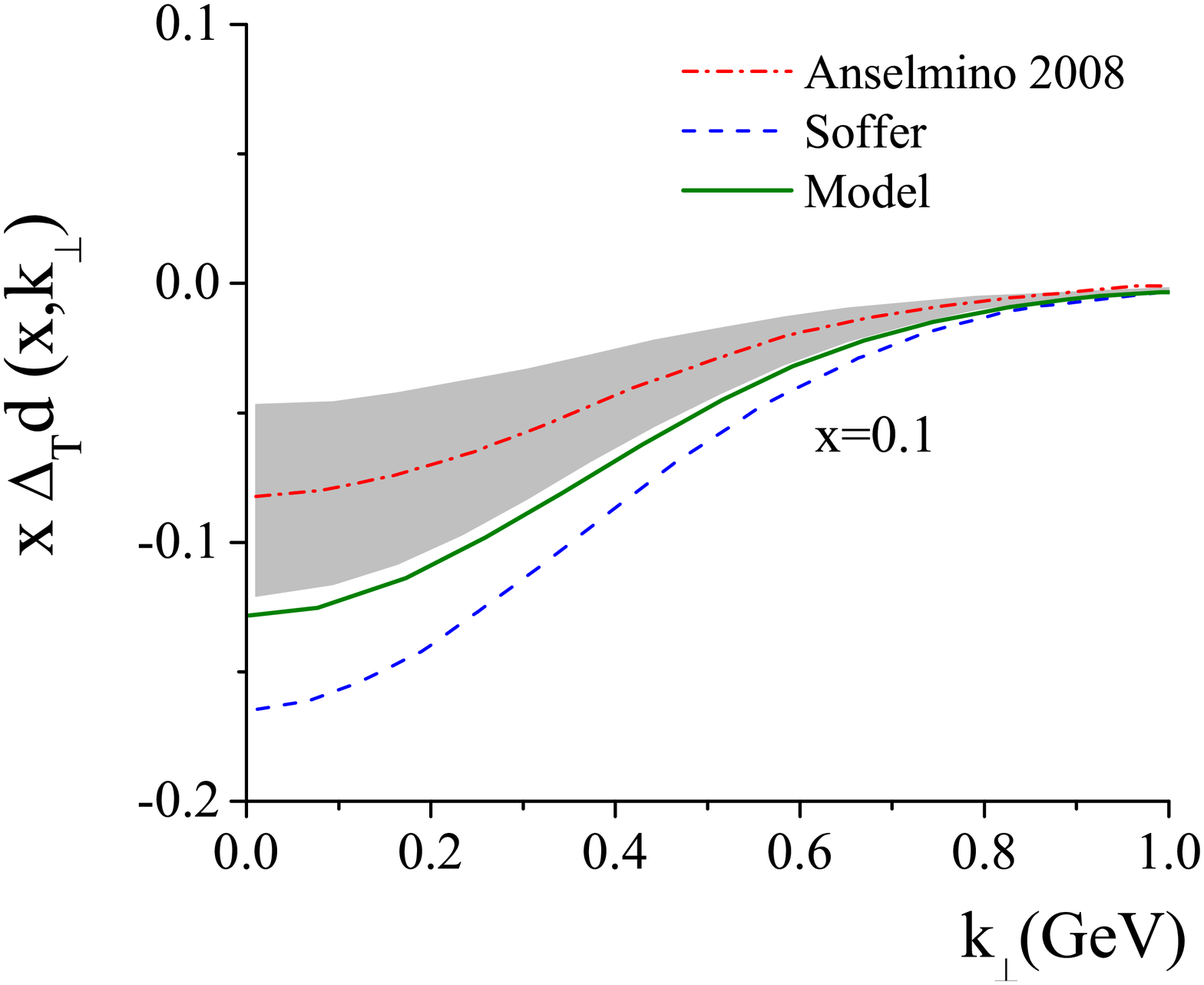}
\end{tabular}}
 \caption{\footnotesize (color online) The transversity distribution function for  valence d  quark calculated by our model
as a function of x and  $k_\bot$  at $Q^{2}=2.4 \ GeV^{2}$ . They are compared with those from Soffer and Anselmino
global fits \cite{Anselmino2009,soffer2}.
 }
\end{figure}

\begin{figure}[htp]\label{xqvt2013}
\centerline{\begin{tabular}{cc}
\includegraphics[width=8.5cm]{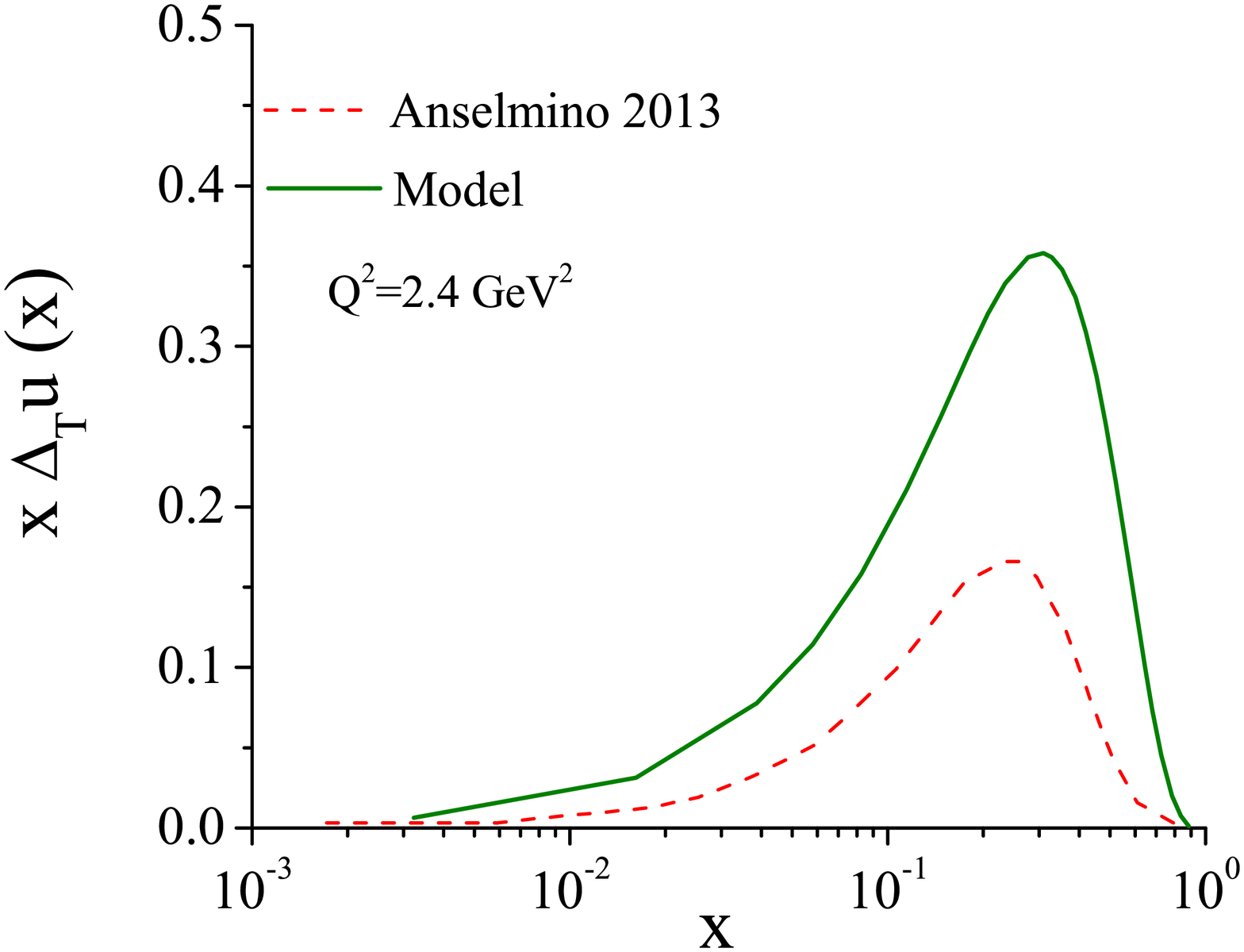}
\includegraphics[width=8.5cm]{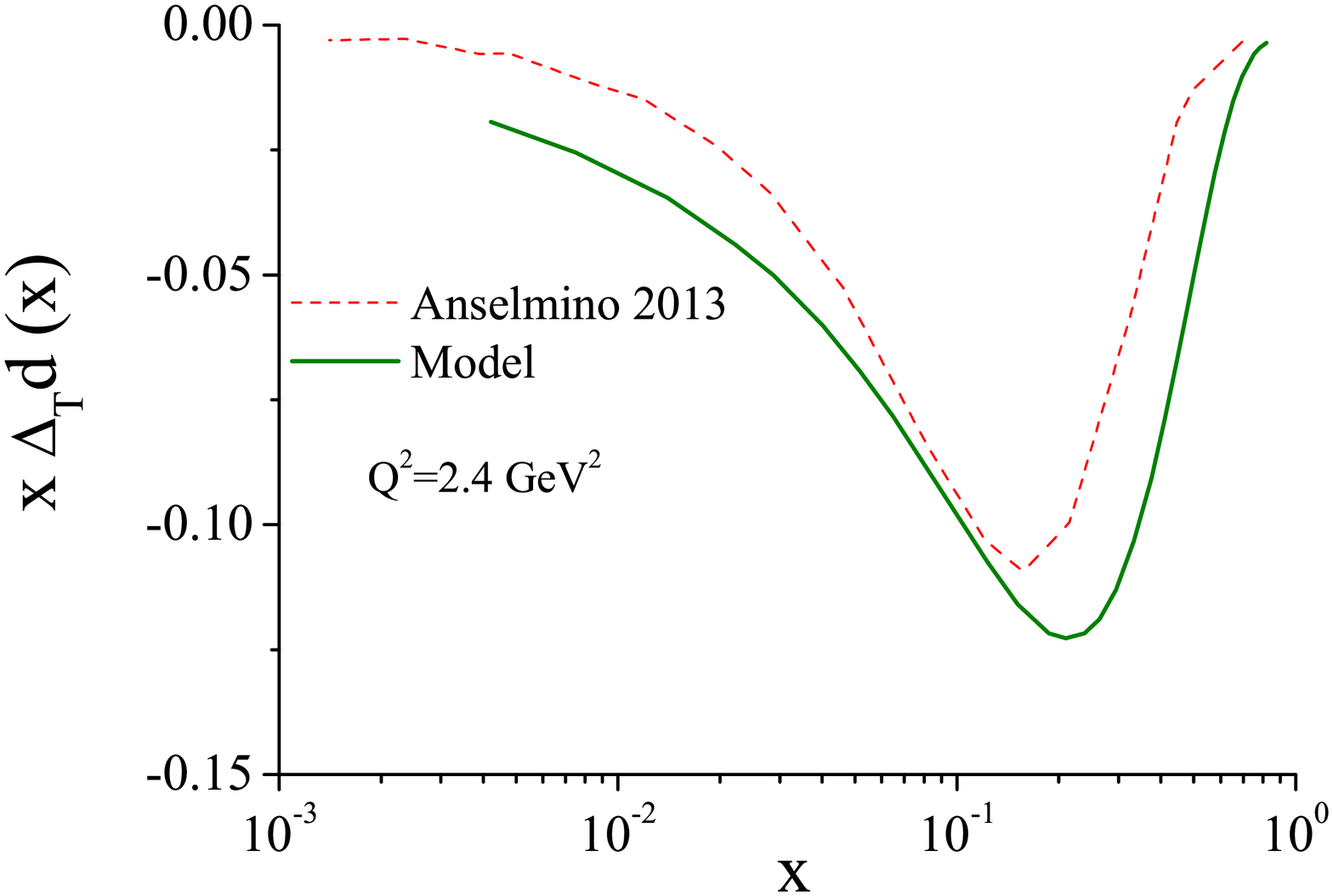}
\end{tabular}}
 \caption{\footnotesize (color online) The transversity distribution functions  for  valence  u and  d  quarks in our
 model  as a function of  x and at $Q^{2}=2.4 \ GeV^{2}$  and  comparison with Anselmino fit (2013)\cite{Anselmino2013}.
 }
\end{figure}
In figure (7) we present  the result for  $  x
[\Delta_{T}u _{v}(x,Q^{2})- \frac{1}{4} \Delta_{T}d _{v} (x,Q^{2})] $ and
compare   with  those reported by HERMES  and COMPASS Collaborations
\cite{Hermes data, Compass data},  as well as  Radici 's model
\cite{Radici} .
\begin{figure}[htp]\label{compare}
\centerline{\begin{tabular}{cc}
\includegraphics[width=8.5cm]{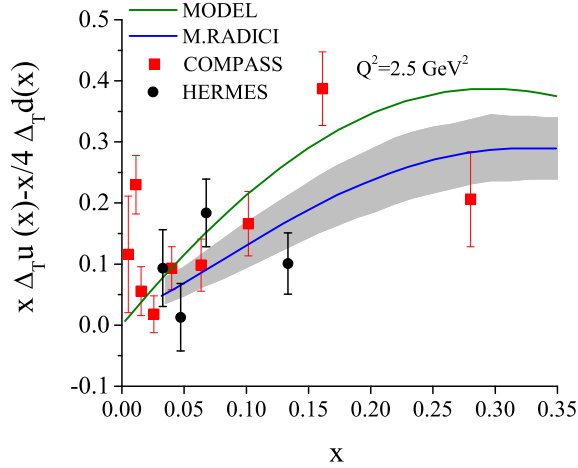}
\end{tabular}}
 \caption{\footnotesize  (color online) The combination of  transversity distribution functions for valence u and d flavors .
 Black circles for SIDIS data from HERMES \cite{Hermes data} , red squares  from CAMPASS \cite{Compass data} and Green curve is obtained by our model at $Q^{2} =2.5 \ GeV^{2}$ .
 Blue curve shows the result of Radici 's model \cite{Radici} with its associated uncertainty
band which represents the same observable as deduced from the
parametrization of Ref\cite{Anselmino band}}.
  \end{figure}
Another interesting quantity, related to the first moment is the
tensor charge,  defined by the integral (\ref{tensor}) as :
\begin{equation}\label{tensor}
\delta q = \int_{0}^{1} dx \ ( \Delta_{T}q-\Delta_{T}\overline{q}
)
\end{equation}
In our analysis the first moment of sea transversity
distributions turn out to be very small: $(-0.00105)$ for $Q^{2}=1
GeV^{2}$. Therefore, the tensor charges are absolutely the first
moment of valence transversity distribution functions. Actually
the valon model predicts that the sea quark polarization are very
small and are consistent with zero. It is undetectable, since  the
valon structure is generated by perturbative dressing in QCD. In
such processes with massless quarks, helicity is conserved and
therefore, the hard gluons can not induce sea quark polarization
perturbatively. The experiments also support this finding
\cite{hermes1,hermes2,compass1,compass2}. Thus we have no sea
polarization in our model.  As a consequence, the first moment of
transversity distributions of u and d quark(Tensor charges ) at
$Q^2=1 GeV^2$ are:
\begin{equation}\label{number}
\delta u = 0.7386    \  , \
 \delta d = -0.3782
\end{equation}
Finally, in figure (8) our results for   tensor charge are
compared with the predictions of some  models \cite{Anselmino2009,
Anselmino2013,number2, number3, number4, number5, number6}.

\begin{figure}[htp]\label{tensorcharge}
\centerline{\begin{tabular}{cc}
\includegraphics[width=8.5cm]{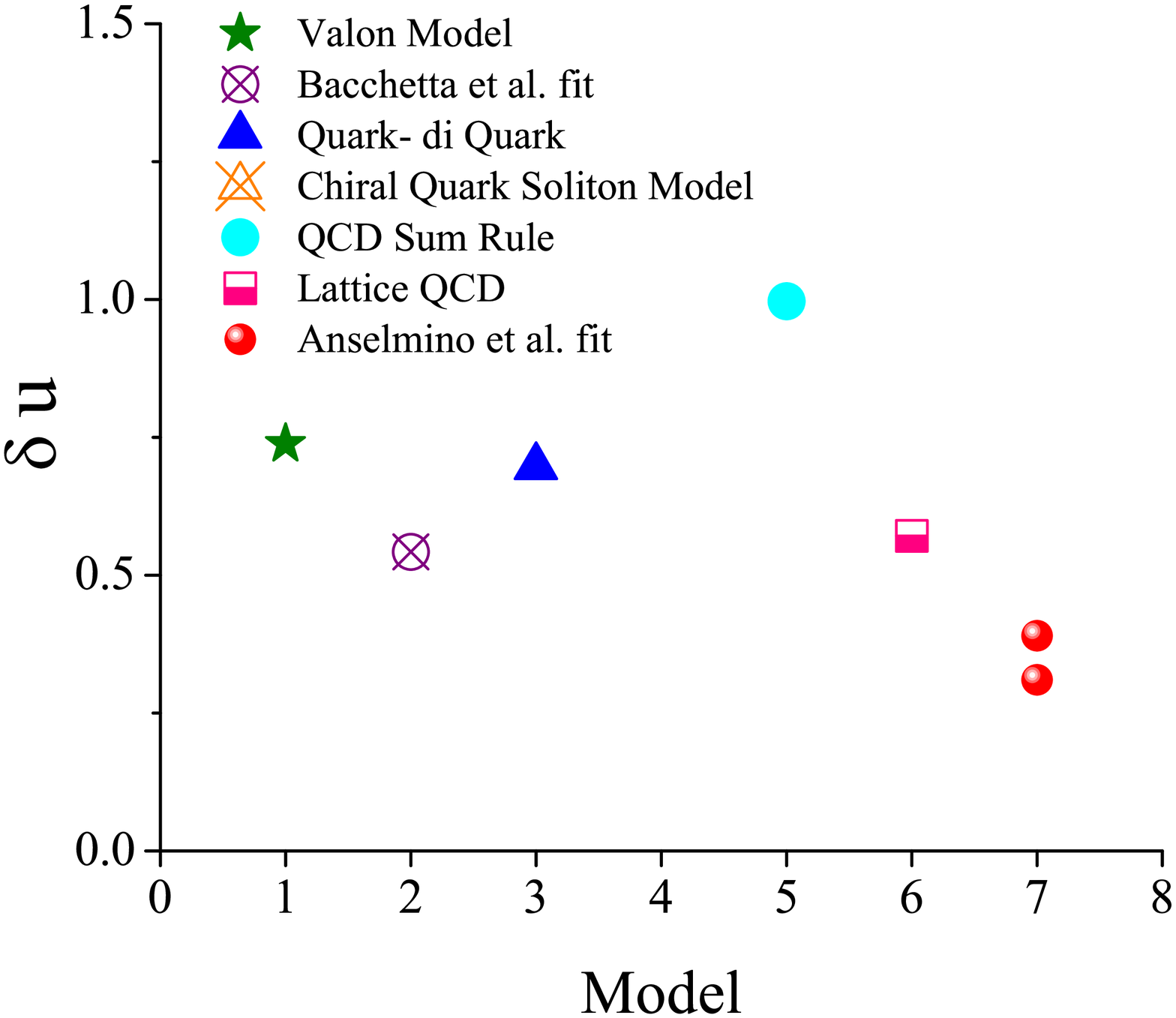}
\includegraphics[width=8.5cm]{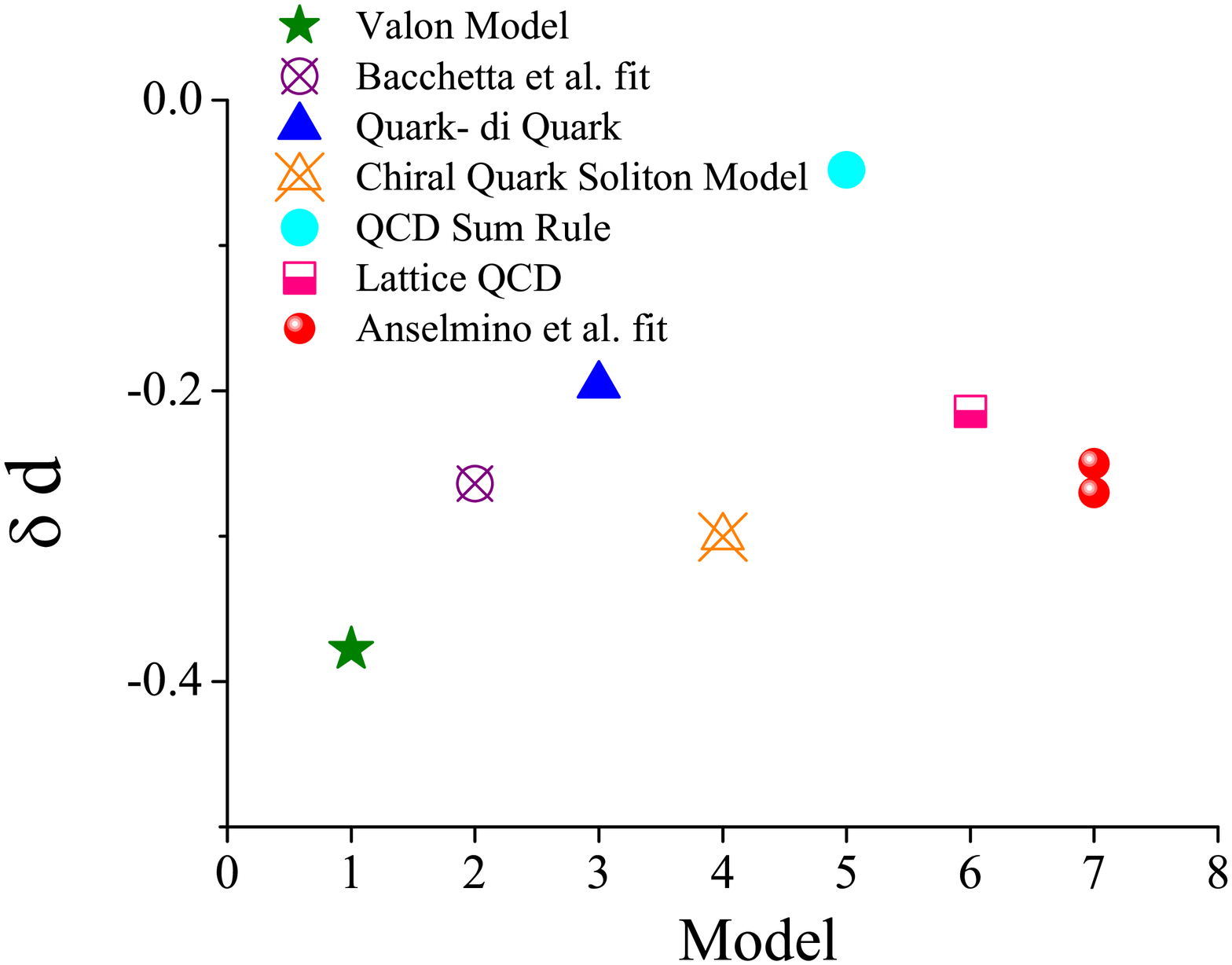}
\end{tabular}}
 \caption{\footnotesize (color online)  The tensor charge for $u$ and   $d$ quarks. Our prediction is shown by number one and  compared with those from several models \cite{Anselmino2013,number2, number3, number4, number5,
 number6}.
 }
\end{figure}

\section{Conclusions and Remarks}
We have utilized the so called valon model and calculated Transversity distribution functions for u and d quarks inside the proton. The transversity distribution functions together with the helicity distribution functions provide a more comprehensive picture of the proton structure. While the former is fairly well understood, the latter is just beginning to be probed. Our calculation in this paper is a step towards this goal. As noted in Eq. (\ref{number}) of the text, in our model the sea partons contribution to the transversity distributions is consistent with zero, whereas the valence sector assumes a sizeable value. In a sense, this prediction is similar to the one we have made for the helicity distribution in Reference \cite{Arash2007}, which later on confirmed by experiment. However, the obtained results do not exhaust the spin of proton and implies that there is room for further contribution from perhaps, the orbital angular momentum. It also shows that a simple model like valon reasonably well reproduces the experimental data and
hence provide a physical picture of the proton structure in the NLO approximation.

\section*{ Acknowledgment}

We would like to thank Professor Mauro Anselmino  for his careful reading of the manuscript
and for the productive discussions.

\section*{Appendix }
Here we list  the anomalous dimensions in mellin space and
$\overline{MS}$ scheme: \cite{Hayashigaki}

 the $ \gamma^{(0)}(n),\gamma^{(1)}(n)$
adequate to $ \Delta_{T}q $ are as follows:
\begin{equation}\label{five}
\Delta_{T}\gamma_{qq}^{(0)}(n)=C_{F}[\frac{3}{2}-2\sum_{j=1}^{n}\frac{1}{j}]
\end{equation}
\begin{eqnarray}
\Delta_{T}\gamma_{qq,\eta}^{(1)}(n) & = & C_{F}^{2}\{\frac{3}{8}+
\frac{2}{n(n+1)}\delta_{\eta-}-3S_{2}(n)-4S_{1}(n)[S_{2}(n)-\acute{S}_{2}(\frac{n}{2})]
\nonumber \\ && -8\tilde{S}(n)+\acute{S}_{3}(\frac{n}{2})\}
\nonumber \\ &&
+\frac{1}{2}C_{F}N_{c}\{\frac{17}{12}-\frac{2}{n(n+1)}\delta_{\eta-}-\frac{134}{9}S_{1}(n)+\frac{22}{3}S_{2}(n)
\nonumber \\ &&
+4S_{1}(n)[2S_{2}(n)-\acute{S}_{2}(\frac{n}{2})]+8\tilde{S}(n)-\acute{S}_{3}(\frac{n}{2})\}
\nonumber \\ &&
+\frac{2}{3}C_{F}T_{F}\{-\frac{1}{4}+\frac{10}{3}S_{1}(n)-2S_{2}(n)\},
\end{eqnarray}
 where $ \eta=\pm $ and the S (Harmonic Functions)
are defined by ;
\begin{eqnarray}
S_k(n) & \equiv & \sum_{j=1}^n \frac{1}{j^k}\\
S_k'\left(\frac{n}{2}\right) & \equiv & 2^{k-1} \sum_{j=1}^n
\frac{1+(-)^j}{j^k} = \frac{1}{2} (1+\eta )
S_k\left(\frac{n}{2}\right)+
\frac{1}{2} (1-\eta ) S_k\left(\frac{n-1}{2}\right)\\
\tilde{S}(n) & \equiv & \sum_{j=1}^n \frac{(-)^j}{j^2} S_1(j)
\nonumber \\ & = & -\frac{5}{8} \zeta (3) +\eta \left[
\frac{S_1(n)}{n^2} + \frac{\pi^2 }{12} G(n) +\int_0^1 dx\; x^{n-1}
\frac{{\rm{Li}}_2(x)} {1+x}\right]
\end{eqnarray}

with $G(n)\equiv \psi\left(\frac{n+1}{2}\right) -
\psi\left(\frac{n}{2}\right)$, $\psi (z)=d\ln \Gamma (z) /dz$ and
$\eta =\pm 1$ for $\delta P_{NS \pm}^{(1)n}$ and $\eta =- 1$ for
the flavor singlet anomalous dimensions.

\end{document}